*Research Note*

# Policy Invariance under Reward Transformations for General-Sum Stochastic Games

**Xiaosong Lu**                                                    LUXIAOS@SCE.CARLETON.CA
**Howard M. Schwartz**                                      SCHWARTZ@SCE.CARLETON.CA
*Department of Systems and Computer Engineering*
*Carleton University*
*1125 Colonel By Drive, Ottawa, ON K1S 5B6 Canada*

**Sidney N. Givigi Jr.**                                      SIDNEY.GIVIGI@RMC.CA
*Department of Electrical and Computer Engineering*
*Royal Military College of Canada*
*13 General Crerar Cres, Kingston, ON K7K 7B4 Canada*

## Abstract

We extend the potential-based shaping method from Markov decision processes to multi-player general-sum stochastic games. We prove that the Nash equilibria in a stochastic game remains unchanged after potential-based shaping is applied to the environment. The property of policy invariance provides a possible way of speeding convergence when learning to play a stochastic game.

## 1. Introduction

In reinforcement learning, one may suffer from the temporal credit assignment problem (Sutton & Barto, 1998) where a reward is received after a sequence of actions. The delayed reward will lead to difficulty in distributing credit or punishment to each action from a long sequence of actions and this will cause the algorithm to learn slowly. An example of this problem can be found in some episodic tasks such as a soccer game where the player is only given credit or punishment after a goal is scored. If the number of states in the soccer game is large, it will take a long time for a player to learn its equilibrium policy.

Reward shaping is a technique to improve the learning performance of a reinforcement learner by introducing shaping rewards to the environment (Gullapalli & Barto, 1992; Mataric, 1994). When the state space is large, the delayed reward will slow down the learning dramatically. To speed up the learning, the learner may apply shaping rewards to the environment as a supplement to the delayed reward. In this way, a reinforcement learning algorithm can improve its learning performance by combining a "good" shaping reward function with the original delayed reward.

The applications of reward shaping can be found in the literature (Gullapalli & Barto, 1992; Dorigo & Colombetti, 1994; Mataric, 1994; Randløv & Alstrøm, 1998). Gullapalli and Barto (1992) demonstrated the application of shaping to a key-press task where a robot was trained to press keys on a keyboard. Dorigo and Colombetti (1994) applied shaping policies for a robot to perform a predefined animate-like behavior. Mataric (1994) presented an intermediate reinforcement function for a group of mobile robots to learn a foraging task. Randløv and Alstrøm (1998) combined reinforcement learning with shaping to make an agent learn to drive a bicycle to a goal. The theoretical analysis of reward shaping can be found in the literature (Ng, Harada, & Russell, 1999; Wiewiora, 2003; Asmuth, Littman, & Zinkov, 2008). Ng et al. (1999) presented a potential-based shaping reward that can guarantee the policy invariance for a single agent in a Markov decision process (MDP). Ng et al. proved that the optimal policy keeps unchanged after adding the potential-based shaping reward to an MDP environment. Following Ng et al., Wiewiora (2003) showed that the effects of potential-based shaping can be achieved by a particular initialization of Q-values for agents using Q-learning. Asmuth et al. (2008) applied the potential-based shaping reward to a model-based reinforcement learning approach.

The above articles focus on applications of reward shaping to a single agent in an MDP. For the applications of reward shaping in general-sum games, Babes, Munoz de Cote, and Littman (2008) introduced a social shaping reward for players to learn their equilibrium policies in the iterated prisoner's dilemma game. But there is no theoretical proof of policy invariance under the reward transformation. In our research, we prove that the Nash equilibria under the potential-based shaping reward transformation (Ng et al., 1999) will also be the Nash equilibria for the original game under the framework of general-sum stochastic games. Note that the similar work of Devlin and Kudenko (2011) was published while this article was under review. But Devlin and Kudenko only proved sufficiency based on a proof technique introduced by Asmuth et al. (2008), while we prove both sufficiency and necessity using a different proof technique in this article.

## 2. Framework of Stochastic Games

Stochastic games were first introduced by Shapley (1953). In a stochastic game, players choose the joint action and move from one state to another state based on the joint action they choose. In this section, under the framework of stochastic games, we introduce Markov decision processes, matrix games and stochastic games respectively.

### 2.1 Markov Decision Processes

A Markov decision process is a tuple $(S, A, T, \gamma, R)$ where $S$ is the state space, $A$ is the action space, $T : S \times A \times S \rightarrow [0, 1]$ is the transition function, $\gamma \in [0, 1]$ is the discount factor and $R : S \times A \times S \rightarrow \mathbb{R}$ is the reward function. The transition function denotes a probability distribution over next states given the current state and action. The reward function denotes the received reward at the next state given the current action and the current state. A Markov decision process has the following Markov property: the player's next state and reward only depend on the player's current state and action. A player's policy $\pi : S \rightarrow A$ is defined as a probability distribution over the player's actions given a state. An optimal policy $\pi^*$ will maximize the player's discounted future reward. For any MDP, there exists a deterministic optimal policy for the player (Bertsekas, 1987).

Starting in the current state $s$ and following the optimal policy thereafter, we can get the optimal state-value function as the expected sum of discounted rewards (Sutton & Barto, 1998)

$$V^{\pi^*}(s) = E\left\{\sum_{j=0}^{T} \gamma^j r_{k+j+1} \big| s_k = s, \pi^*\right\} \tag{1}$$

where $k$ is the current time step, $r_{k+j+1}$ is the received immediate reward at the time step $k+j+1$, $\gamma \in [0, 1]$ is a discount factor, and $T$ is a final time step. In (1), we have $T \rightarrow \infty$ if the task is an infinite-horizon task such that the task will run over infinite period. If the task is episodic, $T$ is defined as the terminal time when each episode is terminated at the time step $T$. Then we call the state where each episode ends as the terminal state $s_T$. In a terminal state, the state-value function is always zero such that $V(s_T) = 0$ for all $s_T \in S$. Given the current state $s$ and action $a$, and following the optimal policy thereafter, we can define an optimal action-value function (Sutton & Barto, 1998)

$$Q^{\pi^*}(s,a) = \sum_{s' \in S} T(s,a,s') \left[ R(s,a,s') + \gamma V^{\pi^*}(s') \right] \tag{2}$$

where $T(s,a,s') = Pr\{s_{k+1} = s' | s_k = s, a_k = a\}$ is the probability of the next state being $s_{k+1} = s'$ given the current state $s_k = s$ and action $a_k = a$ at time step $k$, and $R(s,a,s') = E\{r_{k+1} | s_k = s, a_k = a, s_{k+1} = s'\}$ is the expected immediate reward received at state $s'$ given the current state $s$ and action $a$. In a terminal state, the action-value function is always zero such that $Q(s_T, a) = 0$ for all $s_T \in S$.

## 2.2 Matrix Games

A matrix game is a tuple $(n, A_1, \ldots, A_n, R_1, \ldots, R_n)$ where $n$ is the number of players, $A_i (i = 1, \ldots, n)$ is the action set for the player $i$ and $R_i : A_1 \times \cdots \times A_n \rightarrow \mathbb{R}$ is the payoff function for the player $i$. A matrix game is a game involving multiple players and a single state. Each player $i (i = 1, \ldots, n)$ selects an action from its action set $A_i$ and receives a payoff. The player $i$'s payoff function $R_i$ is determined by all players' joint action from joint action space $A_1 \times \cdots \times A_n$. For a two-player matrix game, we can set up a matrix with each element containing a payoff for each joint action pair. Then the payoff function $R_i$ for player $i (i = 1, 2)$ becomes a matrix. If the two players in the game are fully competitive, we will have a two-player zero-sum matrix game with $R_1 = -R_2$.

In a matrix game, each player tries to maximize its own payoff based on the player's strategy. A player's strategy in a matrix game is a probability distribution over the player's action set. To evaluate a player's strategy, we introduce the following concept of Nash equilibrium. A Nash equilibrium in a matrix game is a collection of all players' policies $(\pi_1^*, \cdots, \pi_n^*)$ such that

$$V_i(\pi_1^*, \cdots, \pi_i^*, \cdots, \pi_n^*) \geq V_i(\pi_1^*, \cdots, \pi_i, \cdots, \pi_n^*), \ \ \forall \pi_i \in \Pi_i, i = 1, \cdots, n \tag{3}$$

where $V_i(\cdot)$ is the expected payoff for player $i$ given all players' current strategies and $\pi_i$ is any strategy of player $i$ from the strategy space $\Pi_i$. In other words, a Nash equilibrium is a collection of strategies for all players such that no player can do better by changing its own strategy given that other players continue playing their Nash equilibrium policies (Başar & Olsder, 1999). We define $Q_i(a_1, \ldots, a_n)$ as the received payoff of the player $i$ given players' joint action $a_1, \ldots, a_n$, and $\pi_i(a_i)$ $(i = 1, \ldots, n)$ as the probability of player $i$ choosing action $a_1$. Then the Nash equilibrium defined in (3) becomes

$$\sum_{a_1, \ldots, a_n \in A_1 \times \cdots \times A_n} Q_i(a_1, \ldots, a_n) \pi_1^*(a_1) \cdots \pi_i^*(a_i) \cdots \pi_n^*(a_n) \geq$$
$$\sum_{a_1, \ldots, a_n \in A_1 \times \cdots \times A_n} Q_i(a_1, \ldots, a_n) \pi_1^*(a_1) \cdots \pi_i(a_i) \cdots \pi_n^*(a_n), \ \ \forall \pi_i \in \Pi_i, i = 1, \cdots, n \tag{4}$$

where $\pi_i^*(a_i)$ is the probability of player $i$ choosing action $a_i$ under the player $i$'s Nash equilibrium strategy $\pi_i^*$.

A two-player matrix game is called a zero-sum game if the two players are fully competitive. In this way, we have $R_1 = -R_2$. A zero-sum game has a unique Nash equilibrium in the sense of the expected payoff. It means that, although each player may have multiple Nash equilibrium strategies in a zero-sum game, the value of the expected payoff $V_i$ under these Nash equilibrium strategies will be the same. If the players in the game are not fully competitive or the summation of the players' payoffs is not zero, the game is called a general-sum game. In a general-sum game, the Nash equilibrium is no longer unique and the game might have multiple Nash equilibria. Unlike the deterministic optimal policy for a single player in an MDP, the equilibrium strategies in a multi-player matrix game may be stochastic.

## 2.3 Stochastic Games

A Markov decision process contains a single player and multiple states while a matrix game contains multiple players and a single state. For a game with more than one player and multiple states, we define a stochastic game (or Markov game) as the combination of Markov decision processes and matrix games. A stochastic game is a tuple $(n, S, A_1, \ldots, A_n, T, \gamma, R_1, \ldots, R_n)$ where $n$ is the number of the players, $T : S \times A_1 \times \cdots \times A_n \times S \to [0, 1]$ is the transition function, $A_i (i = 1, \ldots, n)$ is the action set for the player $i$, $\gamma \in [0, 1]$ is the discount factor and $R_i : S \times A_1 \times \cdots \times A_n \times S \to \mathbb{R}$ is the reward function for player $i$. The transition function in a stochastic game is a probability distribution over next states given the current state and joint action of the players. The reward function $R_i(s, a_1, \ldots, a_n, s')$ denotes the reward received by player $i$ in state $s'$ after taking joint action $(a_1, \ldots, a_n)$ in state $s$. Similar to Markov decision processes, stochastic games also have the Markov property. That is, the player's next state and reward only depend on the current state and all the players' current actions.

To solve a stochastic game, we need to find a policy $\pi_i : S \to A_i$ that can maximize player $i$'s discounted future reward with a discount factor $\gamma$. Similar to matrix games, the player's policy in a stochastic game is probabilistic. An example is the soccer game introduced by Littman (Littman, 1994) where an agent on the offensive side must use a probabilistic policy to pass an unknown defender. In the literature, a solution to a stochastic game can be described as Nash equilibrium strategies in a set of associated state-specific matrix games (Bowling, 2003; Littman, 1994). In these state-specific matrix games, we define the action-value function $Q_i^*(s, a_1, \ldots, a_n)$ as the expected reward for player $i$ when all the players take joint action $a_1, \ldots, a_n$ in state $s$ and follow the Nash equilibrium policies thereafter. If the value of $Q_i^*(s, a_1, \ldots, a_n)$ is known for all the states, we can find player $i$'s Nash equilibrium policy by solving the associated state-specific matrix game (Bowling, 2003). Therefore, for each state $s$, we have a matrix game and we can find the Nash equilibrium strategies in this matrix game. Then the Nash equilibrium policies for the game are the collection of Nash equilibrium strategies in each state-specific matrix game for all the states.

## 2.4 Multi-Player General-Sum Stochastic Games

For a multi-player general-sum stochastic game, we want to find the Nash equilibria in the game if we know the reward function and transition function in the game. A Nash equilibrium in a stochastic game can be described as a tuple of $n$ policies $(\pi_1^*, \ldots, \pi_n^*)$ such that for all $s \in S$ and $i = 1, \cdots, n$,

$$V_i(s, \pi_1^*, \ldots, \pi_i^*, \ldots, \pi_n^*) \geq V_i(s, \pi_1^*, \ldots, \pi_i, \ldots, \pi_n^*) \text{ for all } \pi_i \in \Pi_i \tag{5}$$

where $\Pi_i$ is the set of policies available to player $i$ and $V_i(s, \pi_1^*, \ldots, \pi_n^*)$ is the expected sum of discounted rewards for player $i$ given the current state and all the players' equilibrium policies. To simplify notation, we use $V_i^*(s)$ to represent $V_i(s, \pi_1^*, \cdots, \pi_n^*)$ as the state-value function under Nash equilibrium policies. We can also define the action-value function $Q^*(s, a_1, \cdots, a_n)$ as the expected sum of discounted rewards for player $i$ given the current state and the current joint action of all the players, and following the Nash equilibrium policies thereafter. Then we can get

$$V_i^*(s) = \sum_{a_1,\cdots,a_n \in A_1 \times \cdots \times A_n} Q_i^*(s,a_1,\cdots,a_n)\pi_1^*(s,a_1)\cdots\pi_n^*(s,a_n), \tag{6}$$

$$Q_i^*(s,a_1,\ldots,a_n) = \sum_{s' \in S} T(s,a_1,\ldots,a_n,s')\left[R_i(s,a_1,\ldots,a_n,s') + \gamma V_i^*(s')\right], \tag{7}$$

where $\pi_i^*(s,a_i) \in \mathrm{PD}(A_i)$ is a probability distribution over action $a_i$ under player $i$'s Nash equilibrium policy, $T(s,a_1,\ldots,a_n,s') = Pr\{s_{k+1} = s'|s_k = s,a_1,\ldots,a_n\}$ is the probability of the next state being $s'$ given the current state $s$ and joint action $(a_1,\ldots,a_n)$, and $R_i(s,a_1,\ldots,a_n,s')$ is the expected immediate reward received in state $s'$ given the current state $s$ and joint action $(a_1,\ldots,a_n)$. Based on (6) and (7), the Nash equilibrium in (5) can be rewritten as

$$\sum_{a_1,\ldots,a_n \in A_1 \times \cdots \times A_n} Q_i^*(s,a_1,\ldots,a_n)\pi_1^*(s,a_1)\cdots\pi_i^*(s,a_i)\cdots\pi_n^*(s,a_n) \geq$$
$$\sum_{a_1,\ldots,a_n \in A_1 \times \cdots \times A_n} Q_i^*(s,a_1,\ldots,a_n)\pi_1^*(s,a_1)\cdots\pi_i(s,a_i)\cdots\pi_n^*(s,a_n). \tag{8}$$

## 3. Potential-Based Shaping in General-Sum Stochastic Games

Ng et al. (1999) presented a reward shaping method to deal with the credit assignment problem by adding a potential-based shaping reward to the environment. The combination of the shaping reward with the original reward may improve the learning performance of a reinforcement learning algorithm and speed up the convergence to the optimal policy. The theoretical studies on potential-based shaping methods that appear in the published literature consider the case of a single agent in an MDP (Ng et al., 1999; Wiewiora, 2003; Asmuth et al., 2008). In our research, we extend the potential-based shaping method from Markov decision processes to multi-player stochastic games. We prove that the Nash equilibria under the potential-based shaping reward transformation will be the Nash equilibria for the original game under the framework of general-sum stochastic games.

We define a potential-based shaping reward $F_i(s,s')$ for player $i$ as

$$F_i(s,s') = \gamma\Phi_i(s') - \Phi_i(s), \tag{9}$$

where $\Phi : S \to \mathbb{R}$ is a real-valued shaping function and $\Phi(s_T) = 0$ for any terminal state $s_T$. We define a multi-player stochastic game as a tuple $M = (S,A_1,\ldots,A_n,T,\gamma,R_1,\ldots,R_n)$ where $S$ is a set of states, $A_1,\ldots,A_n$ are players' action sets, $T$ is the transition function, $\gamma$ is the discount factor, and $R_i(s,a_1,\ldots,a_n,s')(i=1,\ldots,n)$ is the reward function for player $i$. After adding the shaping reward function $F_i(s,s')$ to the reward function $R_i(s,a_1,\ldots,a_n,s')$, we define a transformed multi-player stochastic game as a tuple $M' = (S,A_1,\ldots,A_n,T,\gamma,R_1',\ldots,R_n')$ where $R_i'(i=1,\ldots,n)$ is the new reward function given by $R_i'(s,a_1,\ldots,a_n,s') = F_i(s,s') + R_i(s,a_1,\ldots,a_n,s')$. Inspired by Ng et al. (1999)'s proof of policy invariance in an MDP, we prove the policy invariance in a multi-player general-sum stochastic game as follows.

**Theorem 1.** *Given an n-player discounted stochastic game $M = (S,A_1,\ldots,A_n,T,\gamma,R_1,\ldots,R_n)$, we define a transformed n-player discounted stochastic game $M' = (S,A_1,\ldots,A_n,T,\gamma,R_1+F_1,\ldots,R_n+F_n)$ where $F_i \in S \times S$ is a shaping reward function for player i. We call $F_i$ a potential-based shaping function if $F_i$ has the form of (9). Then, the potential-based shaping function $F_i$ is a necessary and sufficient condition to guarantee the Nash equilibrium policy invariance such that*

- *(Sufficiency)* If $F_i$ $(i = 1, \ldots, n)$ is a potential-based shaping function, then every Nash equilibrium policy in $M'$ will also be a Nash equilibrium policy in $M$ (and vice versa).

- *(Necessity)* If $F_i$ $(i = 1, \ldots, n)$ is not a potential-based shaping function, then there may exist a transition function $T$ and reward function $R$ such that the Nash equilibrium policy in $M'$ will not be the Nash equilibrium policy in $M$.

*Proof.* (*Proof of Sufficiency*)

Based on (8), a Nash equilibrium in the stochastic game $M$ can be represented as a set of policies such that for all $i = 1, \ldots, n, s \in S$ and $\pi_{M_i} \in \Pi$

$$\sum_{a_1, \ldots, a_n \in A_1 \times \cdots \times A_n} Q^*_{M_i}(s, a_1, \ldots, a_n) \pi^*_{M_1}(s, a_1) \cdots \pi^*_{M_i}(s, a_i) \cdots \pi^*_{M_n}(s, a_n) \geq$$
$$\sum_{a_1, \ldots, a_n \in A_1 \times \cdots \times A_n} Q^*_{M_i}(s, a_1, \ldots, a_n) \pi^*_{M_1}(s, a_1) \cdots \pi_{M_i}(s, a_i) \cdots \pi^*_{M_n}(s, a_n). \tag{10}$$

We subtract $\Phi_i(s)$ on both sides of (10) and get

$$\sum_{a_1, \ldots, a_n \in A_1 \times \cdots \times A_n} Q^*_{M_i}(s, a_1, \ldots, a_n) \pi^*_{M_1}(s, a_1) \cdots \pi^*_{M_i}(s, a_i) \cdots \pi^*_{M_n}(s, a_n) - \Phi_i(s) \geq$$
$$\sum_{a_1, \ldots, a_n \in A_1 \times \cdots \times A_n} Q^*_{M_i}(s, a_1, \ldots, a_n) \pi^*_{M_1}(s, a_1) \cdots \pi_{M_i}(s, a_i) \cdots \pi^*_{M_n}(s, a_n) - \Phi_i(s). \tag{11}$$

Since $\sum_{a_1, \ldots, a_n \in A_1 \times \cdots \times A_n} \pi^*_{M_1}(s, a_1) \cdots \pi^*_{M_i}(s, a_i) \cdots \pi^*_{M_n}(s, a_n) = 1$, we can get

$$\sum_{a_1, \ldots, a_n \in A_1 \times \cdots \times A_n} [Q^*_{M_i}(s, a_1, \ldots, a_n) - \Phi_i(s)] \pi^*_{M_1}(s, a_1) \cdots \pi^*_{M_i}(s, a_i) \cdots \pi^*_{M_n}(s, a_n) \geq$$
$$\sum_{a_1, \ldots, a_n \in A_1 \times \cdots \times A_n} [Q^*_{M_i}(s, a_1, \ldots, a_n) - \Phi_i(s)] \pi^*_{M_1}(s, a_1) \cdots \pi_{M_i}(s, a_i) \cdots \pi^*_{M_n}(s, a_n). \tag{12}$$

We define

$$\hat{Q}_{M'_i}(s, a_1, \ldots, a_n) = Q^*_{M_i}(s, a_1, \ldots, a_n) - \Phi_i(s). \tag{13}$$

Then we can get

$$\sum_{a_1, \ldots, a_n \in A_1 \times \cdots \times A_n} \hat{Q}_{M'_i}(s, a_1, \ldots, a_n) \pi^*_{M_1}(s, a_1) \cdots \pi^*_{M_i}(s, a_i) \cdots \pi^*_{M_n}(s, a_n) \geq$$
$$\sum_{a_1, \ldots, a_n \in A_1 \times \cdots \times A_n} \hat{Q}_{M'_i}(s, a_1, \ldots, a_n) \pi^*_{M_1}(s, a_1) \cdots \pi_{M_i}(s, a_i) \cdots \pi^*_{M_n}(s, a_n). \tag{14}$$

We now use some algebraic manipulations to rewrite the action-value function under the Nash equilibrium in (7) for player $i$ in the stochastic game $M$ as

$$Q^*_{M_i}(s, a_1, \ldots, a_n) - \Phi_i(s) = \sum_{s' \in S} T(s, a_1, \ldots, a_n, s') \left[ R_{M_i}(s, a_1, \ldots, a_n, s') + \gamma \mathcal{V}^*_{M_i}(s') \right.$$
$$\left. + \gamma \Phi_i(s') - \gamma \Phi_i(s') \right] - \Phi_i(s). \tag{15}$$

Since $\sum_{s' \in S} T(s, a_1, \ldots, a_n, s') = 1$, the above equation becomes

$$Q^*_{M_i}(s, a_1, \ldots, a_n) - \Phi_i(s) = \sum_{s' \in S} T(s, a_1, \ldots, a_n, s') \left[ R_{M_i}(s, a_1, \ldots, a_n, s') \right.$$
$$\left. + \gamma \Phi_i(s') - \Phi_i(s) + \gamma \mathcal{V}^*_{M_i}(s') - \gamma \Phi_i(s') \right]. \tag{16}$$

According to (6), we can rewrite the above equation as

$$Q^*_{M_i}(s, a_1, \ldots, a_n) - \Phi_i(s) = \sum_{s' \in S} T(s, a_1, \ldots, a_n, s') \left[ R_{M_i}(s, a_1, \ldots, a_n, s') + \gamma \Phi_i(s') - \Phi_i(s) \right.$$
$$+ \gamma \sum_{a_1, \ldots, a_n \in A_1 \times \cdots \times A_n} Q^*_{M_i}(s', a'_1, \ldots, a'_n) \pi^*_{M_1}(s', a'_1) \cdots \pi^*_{M_i}(s', a'_n) - \gamma \Phi_i(s') \big]$$
$$= \sum_{s' \in S} T(s, a_1, \ldots, a_n, s') \left\{ R_{M_i}(s, a_1, \ldots, a_n, s') + \gamma \Phi_i(s') - \Phi_i(s) \right.$$
$$+ \gamma \sum_{a_1, \ldots, a_n \in A_1 \times \cdots \times A_n} \left[ Q^*_{M_i}(s', a'_1, \ldots, a'_n) - \Phi_i(s') \right] \pi^*_{M_1}(s', a'_1) \cdots \pi^*_{M_i}(s', a'_n) \Big\}. \quad (17)$$

Based on the definitions of $F_i(s, s')$ in (9) and $\hat{Q}_{M'_i}(s, a_1, \ldots, a_n)$ in (13), the above equation becomes

$$\hat{Q}_{M'_i}(s, a_1, \ldots, a_n) = \sum_{s' \in S} T(s, a_1, \ldots, a_n, s') \left[ R_{M_i}(s, a_1, \ldots, a_n, s') + F_i(s, s') \right.$$
$$+ \gamma \sum_{a_1, \ldots, a_n \in A_1 \times \cdots \times A_n} \hat{Q}_{M'_i}(s', a'_1, \ldots, a'_n) \pi^*_{M_1}(s', a'_1) \cdots \pi^*_{M_i}(s', a'_n) \big]. \quad (18)$$

Since equations (14) and (18) have the same form as equations (6)-(8), we can conclude that $\hat{Q}_{M'_i}(s, a_1, \ldots, a_n)$ is the action-value function under the Nash equilibrium for player $i$ in the stochastic game $M'$. Therefore, we can obtain

$$\hat{Q}_{M'_i}(s, a_1, \ldots, a_n) = Q^*_{M'_i}(s, a_1, \ldots, a_n) = Q^*_{M_i}(s, a_1, \ldots, a_n) - \Phi_i(s). \quad (19)$$

If the state $s$ is the terminal state $s_T$, then we have $\hat{Q}_{M'_i}(s_T, a_1, \ldots, a_n) = Q^*_{M_i}(s_T, a_1, \ldots, a_n) - \Phi_i(s_T) = 0 - 0 = 0$. Based on (14) and $\hat{Q}_{M'_i}(s, a_1, \ldots, a_n) = Q^*_{M'_i}(s, a_1, \ldots, a_n)$, we can find that the Nash equilibrium in $M$ is also the Nash equilibrium in $M'$. Then the state-value function under the Nash equilibrium in the stochastic game $M'$ can be given as

$$V^*_{M'_i}(s) = V^*_{M_i}(s) - \Phi_i(s). \quad (20)$$

(*Proof of Necessity*)

If $F_i$ $(i = 1, \ldots, n)$ is not a potential-based shaping function, we will have $F_i(s, s') \neq \gamma \Phi_i(s') - \Phi_i(s)$. Similar to Ng et al. (1999)'s proof of necessity, we define $\Delta = F_i(s, s') - [\gamma \Phi_i(s') - \Phi_i(s)]$. Then we can build a stochastic game $M$ by giving the following transition function $T$ and player 1's reward function $R_{M_1}(\cdot)$

$$T(s_1, a_1^1, a_2, \ldots, a_n, s_3) = 1,$$
$$T(s_1, a_1^2, a_2, \ldots, a_n, s_2) = 1,$$
$$T(s_2, a_1, \ldots, a_n, s_3) = 1,$$
$$T(s_3, a_1, \ldots, a_n, s_3) = 1,$$
$$R_{M_1}(s_1, a_1, \ldots, a_n, s_3) = \frac{\Delta}{2}, \quad (21)$$
$$R_{M_1}(s_1, a_1, \ldots, a_n, s_2) = 0,$$
$$R_{M_1}(s_2, a_1, \ldots, a_n, s_3) = 0,$$
$$R_{M_1}(s_3, a_1, \ldots, a_n, s_3) = 0,$$

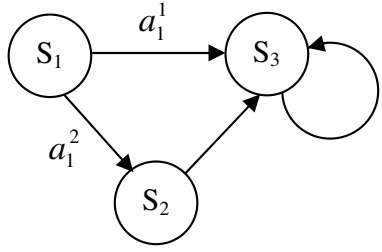

Figure 1: possible states of the stochastic model in the proof of necessity

where $a_i(i = 1, \ldots, n)$ represents any possible action $a_i \in A_i$ from player $i$, and $a_1^1$ and $a_1^2$ represent player 1's action 1 and action 2 respectively. Equation $T(s_1, a_1^1, a_2, \ldots, a_n, s_3) = 1$ in (21) denotes that, given the current state $s_1$, player 1's action $a_1^1$ will lead to the next state $s_3$ no matter what joint action the other players take. Based on the above transition function and reward function, we can get the game model including states $(s_1, s_2, s_3)$ shown in Figure 1. We now define $\Phi_1(s_i) = -F_1(s_i, s_3)(i = 1, 2, 3)$. Based on (6), (7), (19), (20) and (21), we can obtain player 1's action-value function at state $s_1$ in $M$ and $M'$

$$
\begin{aligned}
Q_M^*(s_1, a_1^1, \ldots) &= \frac{\Delta}{2}, \\
Q_M^*(s_1, a_1^2, \ldots) &= 0, \\
Q_{M'}^*(s_1, a_1^1, \ldots) &= F_1(s_1, s_2) + \gamma F_1(s_2, s_3) - \frac{\Delta}{2}, \\
Q_{M'}^*(s_1, a_1^2, \ldots) &= F_1(s_1, s_2) + \gamma F_1(s_2, s_3).
\end{aligned}
$$

Then the Nash equilibrium policy for player 1 at state $s_1$ is

$$
\pi_M^*(s_1, a_1) = \begin{cases} a_1^1 & \text{if } \Delta > 0, \\ a_1^2 & \text{otherwise} \end{cases}, \quad \pi_{M'}^*(s_1, a_1) = \begin{cases} a_1^2 & \text{if } \Delta > 0, \\ a_1^1 & \text{otherwise} \end{cases}. \tag{22}
$$

Therefore, in the above case, the Nash equilibrium policy for player 1 at state $s_1$ in $M$ is not the Nash equilibrium policy in $M'$. $\qquad \square$

The above analysis shows that the potential-based shaping reward with the form of $F_i(s, s') = \gamma \Phi_i(s') - \Phi_i(s)$ guarantees the Nash equilibrium policy invariance. Now the question becomes how to select a shaping function $\Phi_i(s)$ to improve the learning performance of the learner. Ng et al. (1999) showed that $\Phi_i(s) = V_{M_i}^*(s)$ is a good candidate for improving the player's learning performance in an MDP. We substitute $\Phi_i(s) = V^*_{M_i}(s)$ into (18) and get

$$
\begin{aligned}
\hat{Q}_{M'_i}(s, a_1, \ldots, a_n) &= Q^*_{M'_i}(s, a_1, \ldots, a_n) \\
&= \sum_{s' \in S} T(s, a_1, \ldots, a_n, s') \left[ R_{M_i}(s, a_1, \ldots, a_n, s') + F_i(s, s') \right. \\
&+ \gamma \sum_{a_1, \ldots, a_n \in A_1 \times \cdots \times A_n} Q^*_{M'_i}(s', a'_1, \ldots, a'_n) \, \pi^*_{M_1}(s', a'_1) \cdots \pi^*_{M_i}(s', a'_n) ] \\
&= \sum_{s' \in S} T(s, a_1, \ldots, a_n, s') \left[ R_{M_i}(s, a_1, \ldots, a_n, s') + F_i(s, s') \right. \\
&+ \left. \gamma (V^*_{M_i}(s') - \Phi_i(s')) \right] \\
&= \sum_{s' \in S} T(s, a_1, \ldots, a_n, s') \left[ R_{M_i}(s, a_1, \ldots, a_n, s') + F_i(s, s') \right].
\end{aligned}
\tag{23}
$$

Equation (23) shows that the action-value function $Q^*_{M'_i}(s, a_1, \ldots, a_n)$ in state $s$ can be easily obtained by checking the immediate reward $R_{M_i}(s, a_1, \ldots, a_n, s') + F_i(s, s')$ that player $i$ received in state $s'$. However, in practical applications, we will not have all the information of the environment such as $T(s, a_1, \ldots, a_n, s')$ and $R_i(s, a_1, \ldots, a_n, s')$. This means that we cannot find a shaping function $\Phi_i(s)$ such that $\Phi_i(s) = V^*_{M_i}(s)$ without knowing the model of the environment. Therefore, the goal for designing a shaping function is to find a $\Phi_i(s)$ as a "good" approximation to $V^*_{M_i}(s)$.

## 4. Conclusion

A potential-based shaping method can be used to deal with the temporal credit assignment problem and speed up the learning process in MDPs. In this article, we extend the potential-based shaping method to general-sum stochastic games. We prove that the proposed potential-based shaping reward applied to a general-sum stochastic game will not change the original Nash equilibrium of the game. The analysis result in this article has the potential to improve the learning performance of the players in a stochastic game.